# Symmetric *n*-and *p*-Type Sub-5-nm 1D Graphene Nanoribbon Transistors for Homogeneous CMOS Applications

Linqiang Xu,[1,2†] Shiqi Liu,[3†] Qiuhui Li,[1] Ying Li,[1] Shibo Fang,[1] Ying Guo,[4] Yee Sin Ang,[2*], Chen Yang,[1*] Jing Lu,[1, 7, 8, 9*]

[1]State Key Laboratory of Mesoscopic Physics and Department of Physics, Peking University, Beijing 100871, P. R. China

[2]Science, Mathematics and Technology, Singapore University of Technology and Design (SUTD), 8 Somapah Road, Singapore 487372, Singapore

[3]State Key Laboratory of Spintronics Devices and Technologies, Hangzhou 311305, P. R. China

[4]School of Physics and Telecommunication Engineering, Shaanxi Key Laboratory of Catalysis, Shaanxi University of Technology, Hanzhong, 723001, People's Republic of China

[5]Collaborative Innovation Center of Quantum Matter, Beijing 100871, China

[6]Beijing Key Laboratory for Magnetoelectric Materials and Devices, Beijing 100871, China

[7]Peking University Yangtze Delta Institute of Optoelectronics, Nantong 226000, China

[8]Key Laboratory for the Physics and Chemistry of Nanodevices, Peking University, Beijing 100871, China

[*]Corresponding Authors: yeesin_ang@sutd.edu.sg, yangchen96@pku.edu.cn, jinglu@pku.edu.cn

[†]These authors contributed equally to this work

## Abstract

Graphene nanoribbon (GNR) emerges as an exceptionally promising channel candidate due to its tunable sizable bandgap (0-3 eV), ultrahigh carrier mobility (up to 4600 $cm^2$ $V^{-1}$ $s^{-1}$), and excellent device performance (current on-off ratio of $10^7$). However, the asymmetry of reported *n*-type and *p*-type GNR field-effect transistors (FETs) at ultrashort gate length ($L_g$) has become an obstacle to future complementary metal-oxide-semiconductor (CMOS) integration. Here, we conduct *ab initio* quantum transport simulations to investigate the transport properties of sub-5-nm $L_g$ 7 armchair-edge GNR (7 AGNR) FETs. The on-state current, delay time, and power dissipation of the *n*-type and *p*-type 7 AGNR FETs fulfill the International Technology Roadmap for Semiconductors targets for high-performance devices when $L_g$ is reduced to 3 nm. Remarkably, the 7 AGNR FETs exhibit superior *n*-type and *p*-type symmetry to the 7-9-7 AGNR FETs due to the more symmetrical electron/hole effective masses. Compared to the monolayer $MoS_2$ and $MoTe_2$ counterparts, the 7 AGNR FETs have better device performance, which could be further improved via gate engineering. Our results shed light on the immense potential of 7 AGNR in advancing CMOS electronics beyond silicon.

# 1. Introduction

Over the past few decades, the development of integrated circuits is driven by the downscaling of complementary metal-oxide-semiconductor (CMOS) field-effect transistors (FETs).[1, 2] However, with the gate length ($L_g$) reduced to sub-10 nm, traditional silicon (Si)-based FETs have encountered significant short-channel effects (SCEs).[3, 4] Emerging two-dimensional (2D) materials are regarded as a potential substitution of Si due to their excellent gate controllability arising from the atomic thickness and efficient carrier transport induced by the smooth surface.[4-8] As one of the earliest discovered 2D materials, graphene has attracted much attention owing to its extremely high carrier mobility of up to $10^6$ cm$^2$ V$^{-1}$ s$^{-1}$.[9] Nevertheless, the absence of bandgap hinders its further application in electronic devices. Many efforts have been made to open the graphene's bandgap such as adding the vertical electrical field and sandwiching it by boron nitride.[10, 11] The bandgaps opened by these methods are, however, too small to be useful for room temperature device applications. For example, the largest bandgap induced by the vertical electrical field is smaller than 0.3 eV, which is dissatisfactory for the device on-off switch (usually larger than 0.4 eV).[10]

Narrowing 2D materials along one planar direction can effectively increase the bandgap due to the quantum confinement effect.[12] When graphene is reduced to one-dimensional (1D), known as the graphene nanoribbon (GNR), the bandgap as large as 3 eV can be opened.[13] In addition, the carrier mobility of narrow GNR is up to 4600 cm$^2$ V$^{-1}$ s$^{-1}$,[14] which far exceeds many conventional 2D semiconductors like monolayer (ML) $MoS_2$ (200-320 cm$^2$ V$^{-1}$ s$^{-1}$) and $WSe_2$ (250 cm$^2$ V$^{-1}$ s$^{-1}$).[15-17] Experimentally, the grown GNR FETs with narrow widths (< 10 nm) exhibit excellent performance including ultrahigh current on-off ratio of $10^7$ and subthreshold swing below 100 mV/dec.[14, 18] However, the channel lengths of those devices are longer than 100 nm. Recently, the 7-9-7 armchair-edge GNR (7-9-7 AGNR) FETs are predicted to fulfill the International Technology Roadmap for Semiconductors (ITRS) targets at sub-5-nm $L_g$.[19] Although the 7-9-7 AGNR FETs possess outstanding device performance, the biggest challenge arises from the asymmetry of its *n*-type (NMOS) and *p*-type (PMOS) devices, which severely impedes its further application in CMOS integration. Whether symmetric NMOS and PMOS can be achieved in ultrashort GNR FETs remains an open question thus far.

In this work, we propose that 7 AGNR can be an "*np*"-symmetric channel material candidate for sub-5-nm FET device applications.[20] The transport characteristics of the sub-5-nm $L_g$ devices based on 7 AGNR are investigated using *ab initio* quantum transport simulation. Both the *n*-type and *p*-type 7 AGNR FETs can meet the ITRS HP criteria at $L_g$ of 3/5 nm in terms of on-state current, delay time, and power dissipation. The NMOS and PMOS ratios of the above properties for 7 AGNR FETs are all located in 0.5-1.5. Owing to the more symmetrical electron/hole effective masses, 7 AGNR FETs exhibit better NMOS-PMOS symmetry than the 7-9-7 AGNR counterparts. In addition, the overall device performance of 7 AGNR FETs, which can be further enhanced by engineering the gate configurations, outperforms the ML $MoS_2$ and $MoTe_2$ FETs. Hence, our results highlight the strong potential of 7 AGNR for homogeneous CMOS applications beyond silicon.

## 2. Methodology

All calculations were performed in the QuantumATK 2022 software, which combines the density functional theory (DFT) and non-equilibrium Green's function (NEGF) method.[21, 22] The FET usually consists of the (left/right) electrode and the channel regions. The interaction between these two regions can be described by the self-energy $\sum_{k_{//}}^{l/r}$, here $k_{//}$ is the surface-paralleled and transport-direction-perpendicular reciprocal lattice vector. Based on the $\sum_{k_{//}}^{l/r}$, we can obtain the broadening matrix $\Gamma_{k_{//}}^{l/r}(E) = i[\sum_{k_{//}}^{l/r} - (\sum_{k_{//}}^{l/r})^{\dagger}]$ and the retarded central Green's function $G_{k_{//}}(E) = \left[(E + i0^{+})IH_{k_{//}} - \sum_{k_{//}}^{r} - \sum_{k_{//}}^{l}\right]^{-1}$, here $0^{+}$, $I$, and $H_{k_{//}}$ stand for the positive infinitesimal, unit matrix, and channel Hamiltonian, respectively. Herein, the transmission coefficients for every $k_{//}$ vector could be calculated by $T_{k_{//}}(E) = \mathrm{Tr}[\Gamma_{k_{//}}^{l}(E)G_{k_{//}}(E)\Gamma_{k_{//}}^{r}(E)G_{k_{//}}^{\dagger}(E)]$, here $G_{k_{//}}^{\dagger}(E)$ is the advanced central Green's function. Then, we can get the drain current ($I$) according to the Landauer-Büttiker formula:

$$I = \frac{2e}{h}\int_{-\infty}^{+\infty}\{T(E)[f_{S}(E-\mu_{S}) - f_{D}(E-\mu_{D})]\}dE \quad (1)$$

Here, $T(E)$, $f_S$ ($f_D$), and $\mu_S$ ($\mu_D$) stand for the transmission function, Fermi-Dirac distribution of source (drain), and Fermi level of source (drain), respectively. $T(E)$ is calculated by averaging the $T_{k_{//}}(E)$ over $k_{//}$ in the irreducible Brillouin zone. In our simulation, we adopted the double-ξ polarized (DZP) basis set and FHI pseudopotential. The temperature of 300 K is set, and the $k$-point grids of 220×1×1 and 220×1×220 are sampled for the channel and electrode regions, respectively. For the horizontal, vertical, and transport directions, we employ the Neumann, Neumann, and Dirichlet boundary conditions.

The GGA-PBE exchange-correlation function is applied. The DFT-GGA method could offer accurate bandgap estimation in the FET system due to the shielding effects of electron-electron interaction from the doping carriers and dielectric environment.[23-25] Previous studies show that the doping carrier screening results in the bandgap agreement for the degenerately doped ML $MoSe_2$ (1.52/1.59/1.58 eV for GAA/GW/experiment),[26 27, 28] while the dielectric screening leads to the bandgap consistency for the $HfO_2$ sandwiched ML $MoS_2$ (1.76/1.9 eV for GAA/GW).[29, 30] Besides, the feasibility of the DFT-NEGF method in the device simulation has been validated in carbon nanotube FETs at $L_g$ of 5 nm, which exhibits consistent transfer curves and transport properties for simulation and experiment.[31]

## 3. Results

### 3.1 Electronic and Device Configuration

Based on previous studies, zigzag-edge GNRs typically show metallic behavior, whereas armchair-edge GNRs tend to exhibit semiconducting properties.[13] Hence, we choose the armchair-edge GNR as the channel material, whose characteristics are related to the number of carbon atoms ($N$) along the width direction, as depicted in Figure 1(a). Hydrogen atoms are employed to passivate the outermost carbon atoms. Figure 1(b) illustrates the calculated bandgap of armchair-edge GNR for different $N$ based on the DFT-GGA method, here $N$ is selected as $3n$ and $3n+1$ ($n \geq 2$) to include at least two carbon rings. It is observed that the bandgap of GNRs progressively increases as $n$ decreases within each group due to the quantum confinement effect. Generally, a channel material with a larger bandgap is advantageous for suppressing leakage current [$I_{leakage} \propto \exp(-E_g/\alpha k_B T)$, here $k_B$ is the Boltzmann constant, and $T$ is the temperature].[32] Consequently, the 7 AGNR ($N = 7$) that has the largest bandgap of 1.54 eV, which is consistent with the experimental bandgap (1.6 eV),[20] is selected as the channel material. Notably, 7 AGNR has been successfully synthesized experimentally by the bottom-up method.[20] The optimized lattice parameter of 7 AGNR along the $z$ direction is $c = 4.30$ Å, and the electron/hole effective masses ($m_e/m_h$) extracted from the band structure are 0.344 and 0.302 $m_0$, respectively.

To further explore the transport properties, we construct a double-gated (DG) 7 AGNR MOSFET, as depicted in Figure 1(c). The highly doped 7 AGNR is used as the electrode region (including source and drain), while the intrinsic 7 AGNR serves as the channel region. Both the source and drain electrodes are doped with the same carrier type and concentrations. The channel region is divided into two parts, the area covered by the gate and the symmetric underlap (UL) structure between the electrode and gate. A proper UL length is beneficial for improving device performance because the extended total channel length will mitigate the short-channel effect. However, excessively long UL lengths can degrade gate control ability. Thus, selecting a suitable UL length for the 7 AGNR FET is critical. Thus, the UL lengths of 0-3 nm are tested in the simulation. The dielectric region adopts the $SiO_2$ material with a thickness of 0.41 nm at sub-5 nm $L_g$ according to the ITRS requirements.

### 3.2 On-state Current and Subthreshold Swing

The on-state current ($I_{on}$) is a critical figure of merit for evaluating FET performance, with higher values being preferable. $I_{on}$ can be derived from the transfer curves at the on-state point ($V_g^{on}$, $I_{on}$). The on-state voltage $V_g^{on}$ is calculated by the formula $V_g^{on} = V_g^{off} + V_{dd}$, in which $V_g^{off}$ is the off-state voltage, and $V_{dd}$ is the supply voltage. The ITRS 2013 version (denoted as 'ITRS' hereafter), with stricter standards and shorter gate length scaling (down to 5 nm) compared to the latest IRDS 2023 version (down to 12 nm), is used as the evaluation criterion in our simulation.[33, 34] According to the ITRS criteria, $V_{dd}$ is 0.64 V for sub-5 nm $L_g$, and the off-state currents ($I_{off}$) are 0.1 and $5 \times 10^{-5}$ μA/μm for the HP and low-power (LP) devices, respectively. Thus, $V_g^{off}$ can be obtained from the transfer curves at the off-state point ($V_g^{off}$, $I_{off}$), and subsequently, $I_{on}$ can be determined. The doping concentrations ($N_d$) of the 7 AGNR FETs at $L_g = 5$ nm are tested, as displayed in Figure S1. After comparison, $N_d = 5 \times 10^{19}$ cm$^{-3}$ is chosen for both the $n$-type and $p$-type devices.

The transfer characteristics of the $n$-type and $p$-type 7 AGNR FETs at $L_g = 1$, 3, 5 nm are calculated (Figure S2). Since the current is hard to meet the LP $I_{off}$ requirements, we only consider the HP application in the subsequent discussion. HP $I_{on}$ of the 7 AGNR FETs are extracted from the $I$-$V_g$ curves and presented in Table 1. Figures 2(a) and 2(b) display the UL-optimized $I_{on}$ of the $n$-type and $p$-type 7 AGNR FETs, respectively. With UL structure, the ITRS HP requirements for $I_{on}$ are met by both types of devices until $L_g$ down to 3 nm. We also plot the optimal $I_{on}$ of the 7-9-7 AGNR, ML $MoS_2$, ML $MoTe_2$, and 1D Si nanowire (NW) FETs for comparison.[19, 35-37] Overall, both the $n$-type and $p$-type 7 AGNR FETs are inferior to Si NW, while they outperform ML $MoS_2$ and $MoTe_2$ counterparts. Compared to the 7-9-7 AGNR, the 7 AGNR exhibits better $p$- type device performance but poorer $n$-type device performance. These could be explained by the relationship between $I_{on}$ and effective mass $m$, as shown in Figure 2(c). Former studies show that $I_{on}$ reduces with increasing $m$ when $m < 0.45\ m_0$ owing to the decreased carrier velocity, while it increases with enhancing $m$ when $m > 0.45\ m_0$ because of the improved density of states (DOS).[38-40] Both $m_e$ and $m_h$ of the Si NW (0.127 and 0.152 $m_0$) are smaller than those of 7-AGNR (0.344 and 0.302 $m_0$), resulting in overall higher $I_{on}$. The poorer $I_{on}$ performance of ML $MoS_2$ and $MoTe_2$ than 7 AGNR is because their $m_e$ and $m_h$ are very close to 0.45 $m_0$ ($m_e/m_h$ for ML $MoS_2$ and $MoTe_2$ are 0.45/0.45 $m_0$ and 0.56/0.66 $m_0$, respectively). For 7-9-7 AGNR, the smaller electron mass (0.151 $m_0$) than 7 AGNR results in better $n$-type performance, while its higher hole mass (0.354 $m_0$) than 7 AGNR leads to poorer $p$-type performance.

To further evaluate the 7 AGNR FET performance, we extract the subthreshold swing ($SS$) from transfer curves. $SS$ is defined as $SS = \frac{\partial V_g}{\partial \lg I}$ and is used for gate controllability assessment in the subthreshold region. A smaller $SS$ indicates better gate control. Figure 2(d) plots the $SS$ values with and without UL structure for the $n$-type and $p$-type 7 AGNR FETs. The presence of the UL structure significantly reduces $SS$ due to the suppression of the short-channel effect. For the $n$-type and $p$-type devices, $SS$ is decreased by 43/64/81% and 46/66/81% at $L_g$ = 5/3/1 nm, respectively. It is worth noting that the $n$-type and $p$-type devices exhibit good symmetry. To further illustrate the effect of UL structure, we plot the local DOS (LDOS) and spectrum current of the $p$-type 7 AGNR FETs at $L_g$ = 5 nm and $L_{UL}$ = 0/1 nm. The hole barrier height, which is labeled as $\Phi_B$, is calculated by the difference between the Fermi level of the source ($\mu_s$) and the valance band maximum (VBM) of the central channel. On the other hand, the spectrum current consists of the tunneling current ($I_{tunnel}$) and thermionic current ($I_{therm}$).[41, 42]

At the off-state, the $I_{off}$ value is fixed at 0.1 μA/μm for different UL structure devices. To achieve the same $I_{off}$, the 1-nm-UL 7 AGNR FET should possess a smaller barrier height (that is smaller $\Phi_B$ of 0.39 eV) than that of 0-nm-UL 7 AGNR FET ($\Phi_B$ of 0.41 eV) because the barrier width ($w$) is larger for longer UL length. The smaller $\Phi_B$ further leads to the presence of $I_{therm}$ for 1-nm-UL 7 AGNR FET. When the gate voltage of 0.64 V is applied to the device, the 7 AGNR FET is switched from the off-state to the on-state. The $\Phi_B$ is reduced from 0.41 and 0.39 eV to 0 and -0.09 eV for $L_{UL}$ = 0 and 1 nm, respectively. The variation of $\Phi_B$ for 1-nm-UL 7 AGNR FET (0.48 eV) is larger than that of the 0-nm-UL counterpart (0.41 eV), indicating better gate control for longer UL devices. Besides, $I_{tunnel}$ has vanished, and the peak $I_{therm}$ of $L_{UL}$ = 1 nm is about one order of magnitude than that of $L_{UL}$ = 0 nm. The UL structure thus improves both gate controllability and on-state current by modulating $\Phi_B$.

### 3.3 Delay Time, Power Dissipation, and Energy-delay Product

To assess the operating speed of a transistor, the commonly used parameter is the intrinsic delay time ($\tau$), calculated by $\tau = C_t V_{dd}/I_{on}$. $C_t$ represents the total capacitance, which is three times the intrinsic gate capacitance ($C_g$) according to the ITRS criteria. Here $C_g$ is defined as the derivative of the total charge in the channel region ($Q_{ch}$) with respect to $V_g$, given by $C_g = \partial Q_{ch}/\partial V_g$. The $C_t$ values of the 7 AGNR FETs are shown in Table 1. A lower $\tau$ indicates faster operating speed. Figures 4(a) and 4(b) illustrate the UL-optimized $\tau$ of the 7 AGNR, 7-9-7 AGNR, ML $MoS_2$, ML $MoTe_2$, and Si NW FETs.[19, 35-37] When $L_g$ is scaled down to 1 nm, $\tau$ of both $n$-type and $p$-type 7 AGNR FETs can reach ITRS HP targets. Compared to the ML $MoS_2$ and $MoTe_2$ counterparts, 7 AGNR FETs exhibit lower $\tau$ for both types of devices due to the larger $I_{on}$ and smaller $C_t$ at almost all $L_g$ (except for the $n$-type device at $L_g$ = 1 nm). Additionally, $\tau$ of $n$-type and $p$-type 7 AGNR FETs are comparable with Si NW FETs at $L_g$ = 3 and 5 nm, but superior at $L_g$ = 1 nm. Owing to the higher $I_{on}$ and lower $V_{dd}$, 7-9-7 AGNR FETs possess smaller $\tau$ for all the devices than 7 AGNR FETs.

Another major concern for a transistor is power dissipation (PDP), defined as PDP = $V_{dd}I_{on}\tau = C_tV_{dd}^2$. A lower PDP indicates less energy consumption. Figures 4(c) and 4(d) depict the optimal PDP of the 7 AGNR, 7-9-7 AGNR, ML $MoS_2$, ML $MoTe_2$, and Si NW FETs.[19, 35-37] The scaling limit of both $n$-type and $p$-type 7 AGNR FETs can reach 1 nm based on the ITRS targets. For the $n$-type devices, the 7 AGNR FETs exhibit lower PDP than the ML $MoS_2$, ML $MoTe_2$, and Si NW FETs because of the smaller $C_t$ at all gate lengths. The smaller $V_{dd}$ of 7-9-7 AGNR FETs leads to lower PDP than 7 AGNR FETs. As to the $p$-type devices, PDPs of the 7 AGNR FETs are superior to ML $MoTe_2$ and Si NW counterparts but inferior to the 7-9-7 AGNR counterparts. The PDP values of the 7 AGNR FETs are comparable with those of the ML $MoS_2$ FETs. Taking both the switching speed and energy consumption into consideration, we can obtain the energy-delay product (EDP) by EDP = $\tau \times$ PDP. The comparison of EDP at $L_g$ of 3 nm between 7 AGNR and other low-dimensional material FETs is plotted in Figure 6.[19, 35-37, 43, 44] We find that EDPs of the 7 AGNR FETs are superior to the ML $MoS_2$, ML $MoTe_2$, Si Fin, Si NW, InP NW counterparts. Therefore, these results shed light on the good performance of 7 AGNR FETs compared to the other low-dimensional material devices in terms of $\tau$, PDP, and EDP.

### 3.4 Symmetric *n*- and *p*-Type Performance

To achieve better logic operation in CMOS, the symmetric $n$-type and $p$-type characteristics are necessary, especially using the same channel material. Either worse NMOS or PMOS properties will result in a weaker output “high” (that is “1”) or “low” (that is “0”) state. For 7 AGNR FETs, the UL-optimized figure of merits (including $I_{on}$, $SS$, $\tau$, PDP) of the $n$-type and $p$-type devices are depicted in Figures 6(a)-6(d). We observe that the variation trends of $n$-type and $p$-type 7 AGNR FETs are highly consistent for all those properties. This could be further quantified by the NMOS/PMOS ratio, as plotted in Figure 6(e). The ratios of $I_{on}$, $SS$, $\tau$, and PDP all fall within the range of 0.5-1.5. In particular, the $SS$ ratios are smaller than 1.14 for all $L_g$, implying excellent NMOS and PMOS symmetry. Previous study predicted the InAs nanowire FETs exhibit good symmetry at different diameters.[45] After comparison, we find that the $I_{on}$ ratios of 7 AGNR FETs (0.637-0.749) are superior to the InAs FETs at the diameter of 1.6 nm (0.232) and 0.8 nm (0.607).

The symmetric $n$-type and $p$-type behavior of 7 AGNR FETs is attributed to the symmetric electron (0.344 $m_0$) and hole (0.302 $m_0$) effective masses ($m_e/m_h$ = 1.14). Although 7-9-7 AGNR FETs also display excellent device performance, even superior to 7 AGNR FETs, the asymmetry in their NMOS or PMOS characteristics poses a significant obstacle for CMOS applications. Taking on-state current as an example, we compare the NMOS/PMOS ratios between 7 AGNR and 7-9-7 AGNR FETs in Figure 6(f).[19] For all the gate lengths, the ratios of 7 AGNR FETs are closer to 1 than the 7-9-7-AGNR counterparts. This is because the $m_e$ (0.151 $m_0$) and $m_h$ (0.354 $m_0$) of 7-9-7 AGNR exhibit a large discrepancy ($m_e/m_h = 0.43$).[19] Therefore, we can conclude that 7-AGNR is more suitable than 7-9-7-AGNR for homogeneous CMOS integration.

## 4. Discussion

In the current Si FET-based industry, gate engineering from DG to Fin to gate-all-around (GAA) structure has been proven to be an effective method for boosting device performance. From the perspective of natural length λ, the gate numbers ($N$) increase in order (DG to Fin to GAA), resulting in smaller λ ($\lambda \propto 1/N$) and thus better gate control.[4, 46] Inspired by the positive effect of gate engineering, we applied the Fin and GAA structure to both the $n$-type and $p$-type 7 AGNR FETs at $L_g$ = 5 nm and $L_{UL}$ = 2 nm (Figure 7(a)). The variation of $I_{on}$, $\tau$, and PDP are displayed in Figures 7(b)-7(d). $I_{on}$ increases with increasing gate numbers for both types of devices, with improvement ratios reaching 29/66% ($n$-type) and 57/62% ($p$-type) for Fin/GAA structures. Since $\tau$ is inversely proportional to $I_{on}$, the increase of $I_{on}$ leads to the reduction of $\tau$. For Fin/GAA structures, $\tau$ is decreased by 21/35% ($n$-type) and 32/38% ($p$-type). Although PDP is enhanced due to the increased $C_t$, the changes are all smaller than 10%, indicating minimal variation. Consequently, these results suggest that the performance of 7 AGNR FET could be improved by modulating the gate numbers.

## 5. Conclusion

In conclusion, we simulate the transport characteristics of sub-5-nm $L_g$ 7 AGNR FETs based on the first-principles quantum transport simulation. With the assistance of UL structure, $I_{on}$, $\tau$, and PDP of both the $n$-type and $p$-type 7 AGNR FETs at $L_g$ = 3 and 5 nm can meet the HP demands outlined by ITRS. Owing to the symmetric $m_e$ and $m_h$, the 7 AGNR exhibit superior NMOS and PMOS symmetry compared to the 7-9-7 AGNR, with $I_{on}$ ratio closer to 1. Finally, we find that the overall performance of the 7-AGNR FETs is better than the ML $MoS_2$ and $MoTe_2$ counterparts, and it could be further improved through gate engineering. Hence, our findings suggest the 7 AGNR is a promising channel candidate for next-generation CMOS integration.

### Data Availability Statement

The data that support the findings of this study are available from the corresponding authors upon reasonable request.

### Conflict of Interest

The authors declare no conflict of interest.

### Acknowledgment

This work is supported by the Ministry of Science and Technology of China (No.2022YFA1203904 and No.

2022YFA1200072), the National Natural Science Foundation of China (No. 91964101, No. 12274002, and No. 12164036, No. 62174074), the China Scholarship Council, the Fundamental Research Funds for the Central Universities, and the High-performance Computing Platform of Peking University. Y.S.A. and L.X acknowledge the support of Singapore University of Technology Start-up Research Grant (SRG) under the award number SRG SCI 2021 163. Y.S.A. is also supported by SMU-SUTD Joint Grant (Award No. 22-SIS-SMU-054) and SUTD-ZJU Thematic Research Grant (Award No. SUTD-ZJU (TR) 202203). The computational work for this article was partially performed on resources of the National Supercomputing Centre, Singapore (https://www.nscc.sg).

Table 1. Benchmark of the calculated ballistic performance of the *n*- and *p*-type 7 AGNR FETs against the ITRS 2013 HP requirements on the 2028 horizon. The $I_{off}$ value is set as 0.1 μA/μm according to the ITRS criteria.

| | $L_g$ (nm) | $L_{UL}$ (nm) | $I_{on}$ (μA/μm) | $I_{on}/I_{off}$ | SS (mV/dec) | $C_t$ (fF/μm) | $\tau$ (ps) | PDP (fJ/μm) |
|---|---|---|---|---|---|---|---|---|
| ***n*-type HP** | 5 | 0 | 997 | $9.97\times10^3$ | 136 | 0.547 | 0.351 | 0.224 |
| | | 1 | 2674 | $2.67\times10^4$ | 104 | 0.426 | 0.102 | 0.174 |
| | | 2 | 1022 | $1.02\times10^4$ | 84 | 0.270 | 0.168 | 0.110 |
| | | 3 | 430 | $4.30\times10^3$ | 77 | 0.200 | 0.297 | 0.082 |
| | 3 | 0 | - | - | 300 | - | - | - |
| | | 1 | 67 | $6.70\times10^2$ | 179 | 0.178 | 1.704 | 0.073 |
| | | 2 | 905 | $9.05\times10^3$ | 122 | 0.188 | 0.133 | 0.077 |
| | | 3 | 430 | $4.30\times10^3$ | 109 | 0.127 | 0.188 | 0.052 |
| | 1 | 0 | - | - | 870 | - | - | - |
| | | 1 | - | - | 435 | - | - | - |
| | | 2 | 28 | $2.80\times10^2$ | 216 | 0.077 | 1.760 | 0.032 |
| | | 3 | 193 | $1.93\times10^3$ | 169 | 0.064 | 0.213 | 0.026 |
| ***p*-type HP** | 5 | 0 | 2168 | $2.17\times10^4$ | 126 | 0.550 | 0.162 | 0.225 |
| | | 1 | 3580 | $3.58\times10^4$ | 97 | 0.436 | 0.078 | 0.179 |
| | | 2 | 1547 | $1.55\times10^4$ | 80 | 0.302 | 0.125 | 0.124 |
| | | 3 | 963 | $9.63\times10^3$ | 68 | 0.226 | 0.150 | 0.093 |
| | 3 | 0 | - | - | 285 | - | - | - |
| | | 1 | 260 | $2.60\times10^3$ | 174 | 0.234 | 0.576 | 0.096 |
| | | 2 | 1209 | $1.21\times10^4$ | 116 | 0.188 | 0.100 | 0.077 |
| | | 3 | 821 | $8.21\times10^3$ | 98 | 0.154 | 0.120 | 0.063 |
| | 1 | 0 | - | - | 826 | - | - | - |
| | | 1 | - | - | 427 | - | - | - |
| | | 2 | 22 | $2.20\times10^2$ | 216 | 0.082 | 2.380 | 0.034 |
| | | 3 | 303 | $3.03\times10^3$ | 160 | 0.076 | 0.161 | 0.031 |
| **ITRS 2028 horizon** | **5.1** | **-** | **900** | $\mathbf{9.00\times10^3}$ | **-** | **0.6** | **0.423** | **0.24** |

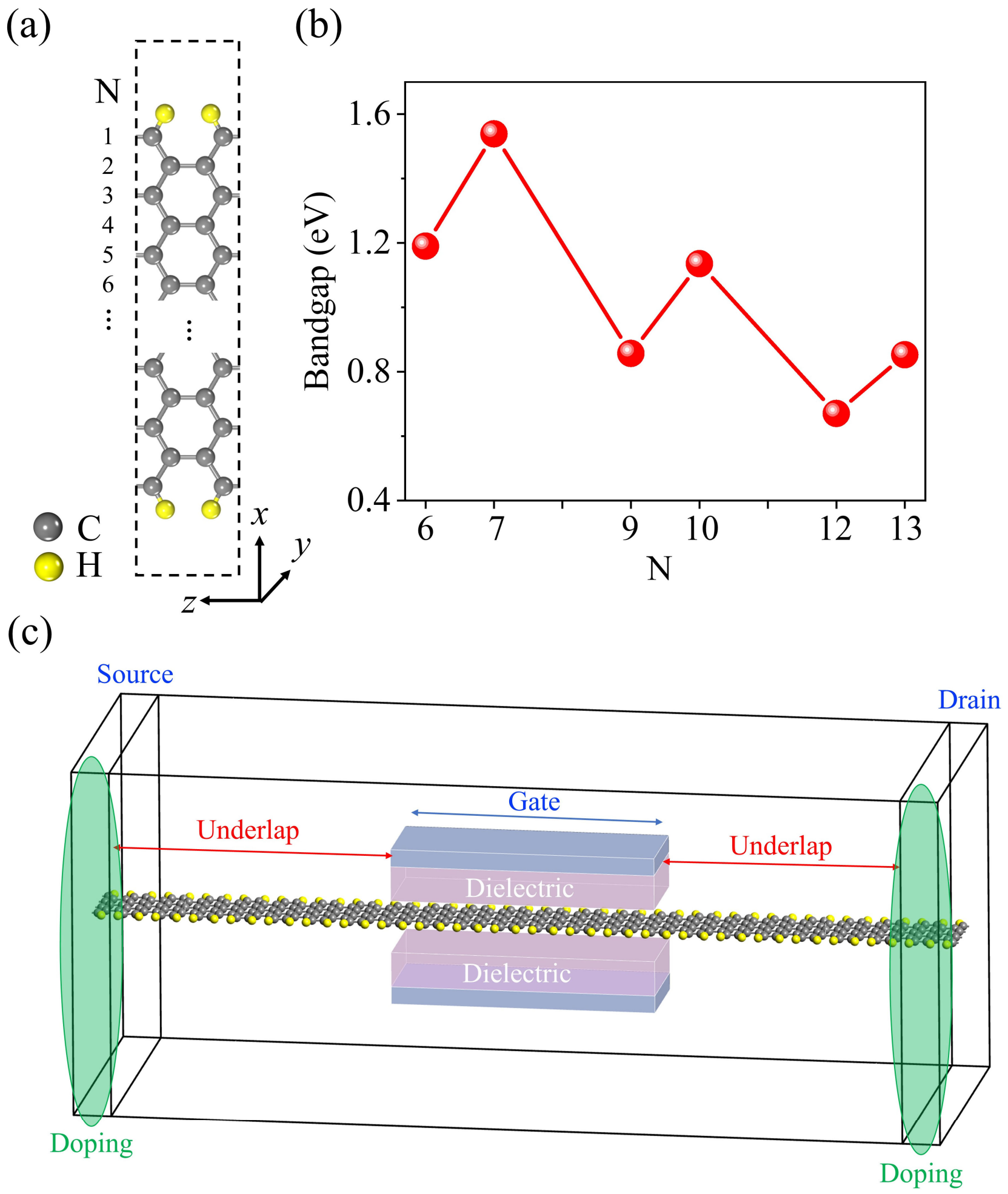

**Figure 1.** (a) Top-view atomic structures of AGNR. *N* indicates the number of carbon atoms along the width direction (*x* direction). (b) Bandgap of the AGNR with respect to the width index *N*. (c) Schematic diagram of the DG 7 AGNR FET. Green region: the doped electrodes.

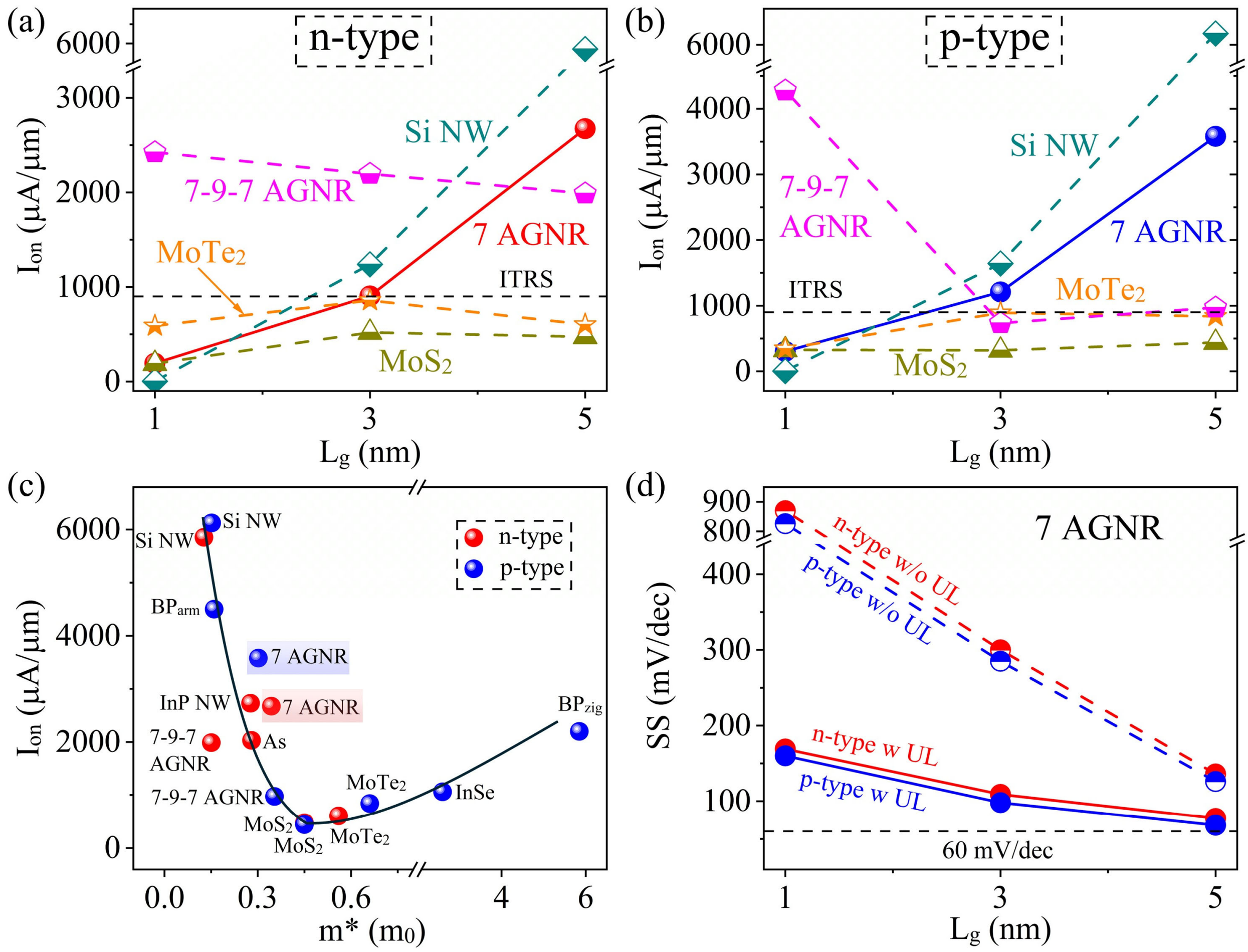

**Figure 2.** UL-optimized $I_{on}$ of the 7 AGNR, 7-9-7 AGNR, Si NW, ML $MoS_2$, and ML $MoTe_2$ FETs versus $L_g$ for the (a) *n*-type and (b) *p*-type devices. The black dashed lines indicate the ITRS HP criteria. (c) HP $I_{on}$ of the low-dimensional material FETs at $L_g$ = 5 nm against $m^*$ of the low-dimensional materials. The black line is the guide to the eyes. (d) Gate length scaling of *SS* in the 7 AGNR FETs for different types. Both the devices with UL (w UL) and without UL (w/o UL) are displayed.

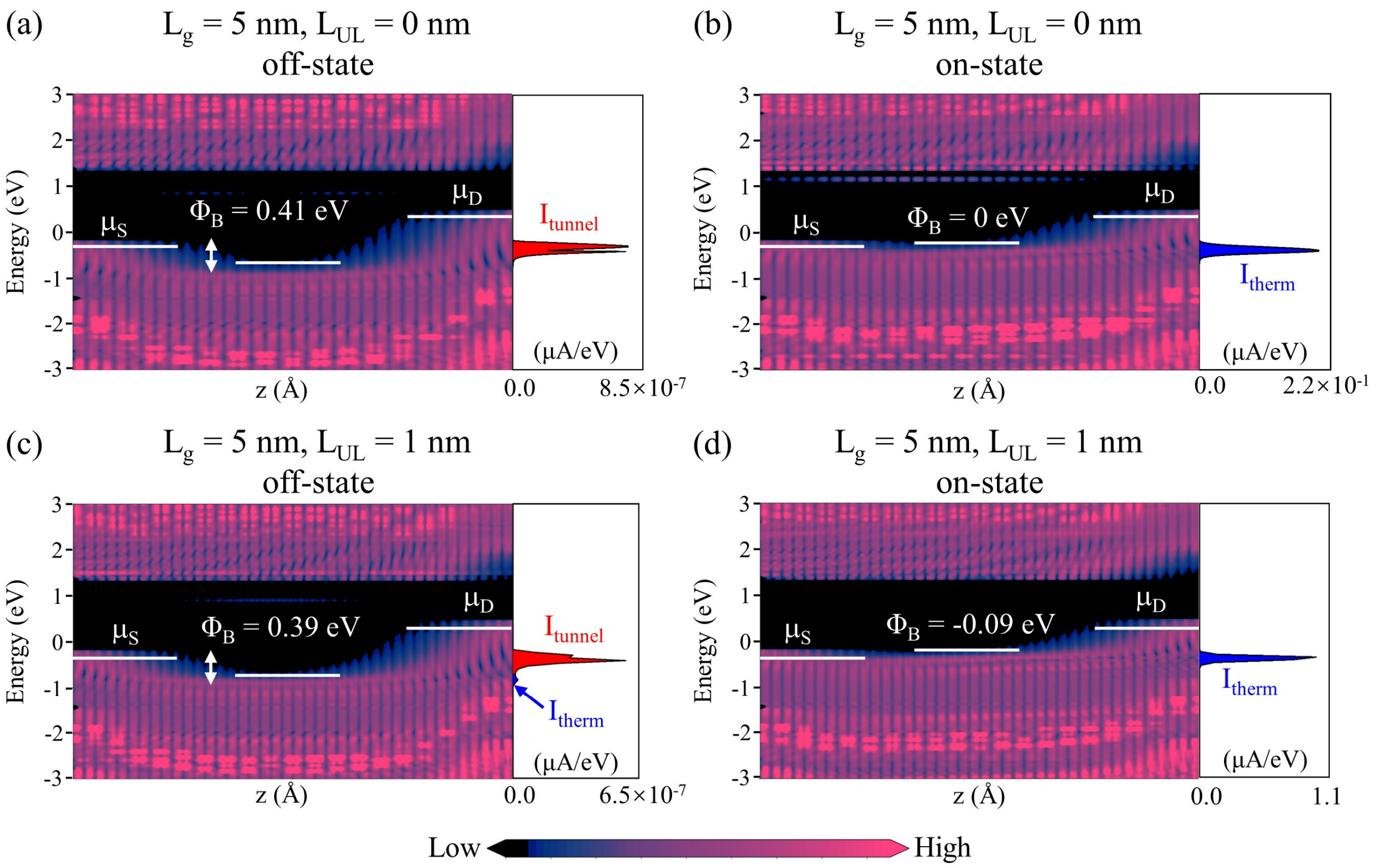

**Figure 3.** LDOS and spectrum current of 5-nm-$L_g$ 7 AGNR FETs at (a) off-state with $L_{UL}$ = 0 nm, (c) on-state with $L_{UL}$ = 0 nm, (c) off-state with $L_{UL}$ = 1 nm, and (d) on-state with $L_{UL}$ = 1 nm. $\Phi_B$ depicts the effective barrier height for hole transportation from source to drain. $\mu_S$ and $\mu_D$ are the electrochemical potentials of the source and train, respectively.

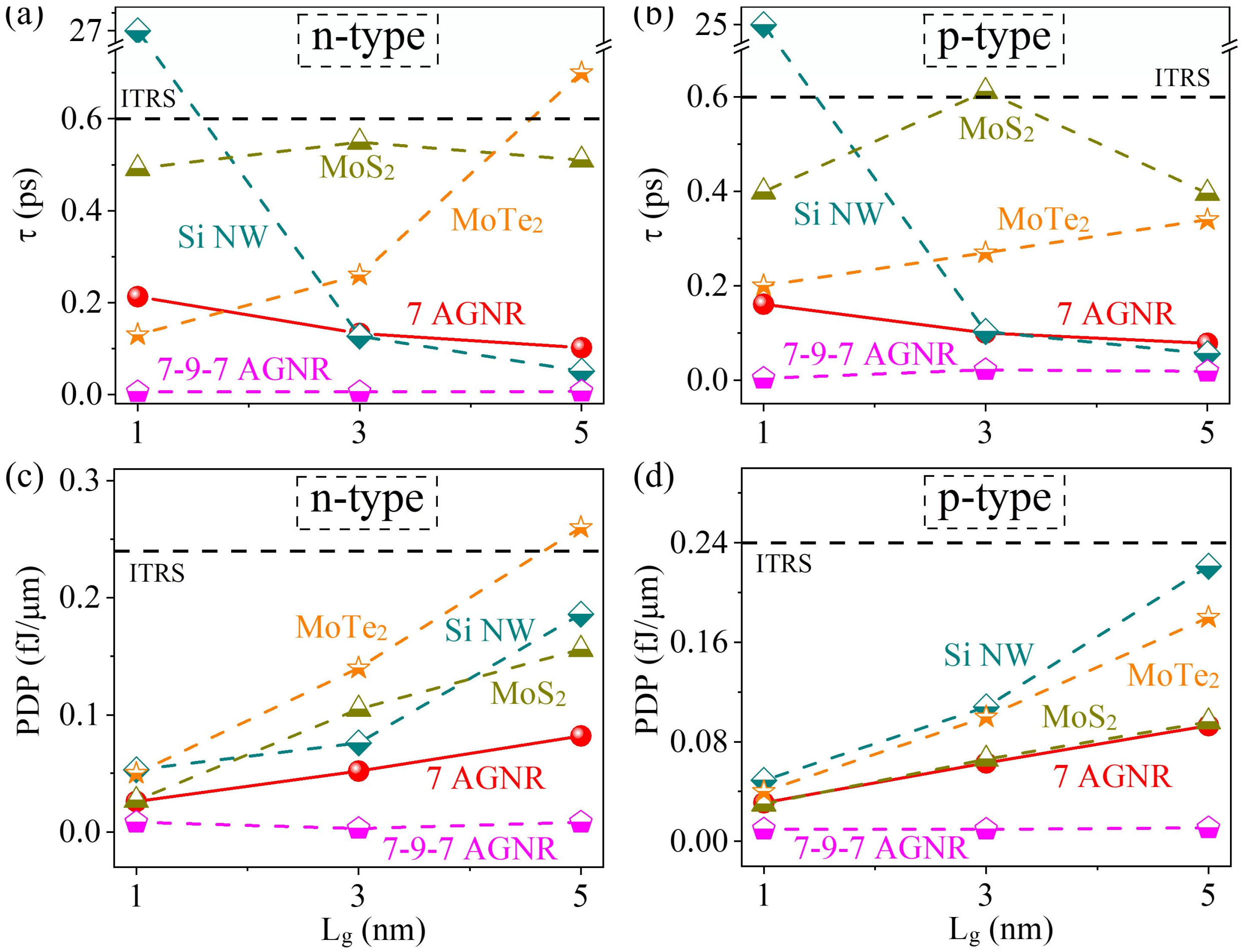

**Figure 4.** UL-optimized (a) $\tau$ of the *n*-type, (b) $\tau$ of the *p*-type, (c) PDP of the *n*-type, and (d) PDP of the *p*-type devices as a function of $L_g$ for the 7 AGNR (red), 7-9-7 AGNR (magenta), Si NW (dark cyan), ML $MoS_2$ (dark yellow), and ML $MoTe_2$ (orange) FETs. The black dashed line indicates the ITRS HP criteria.

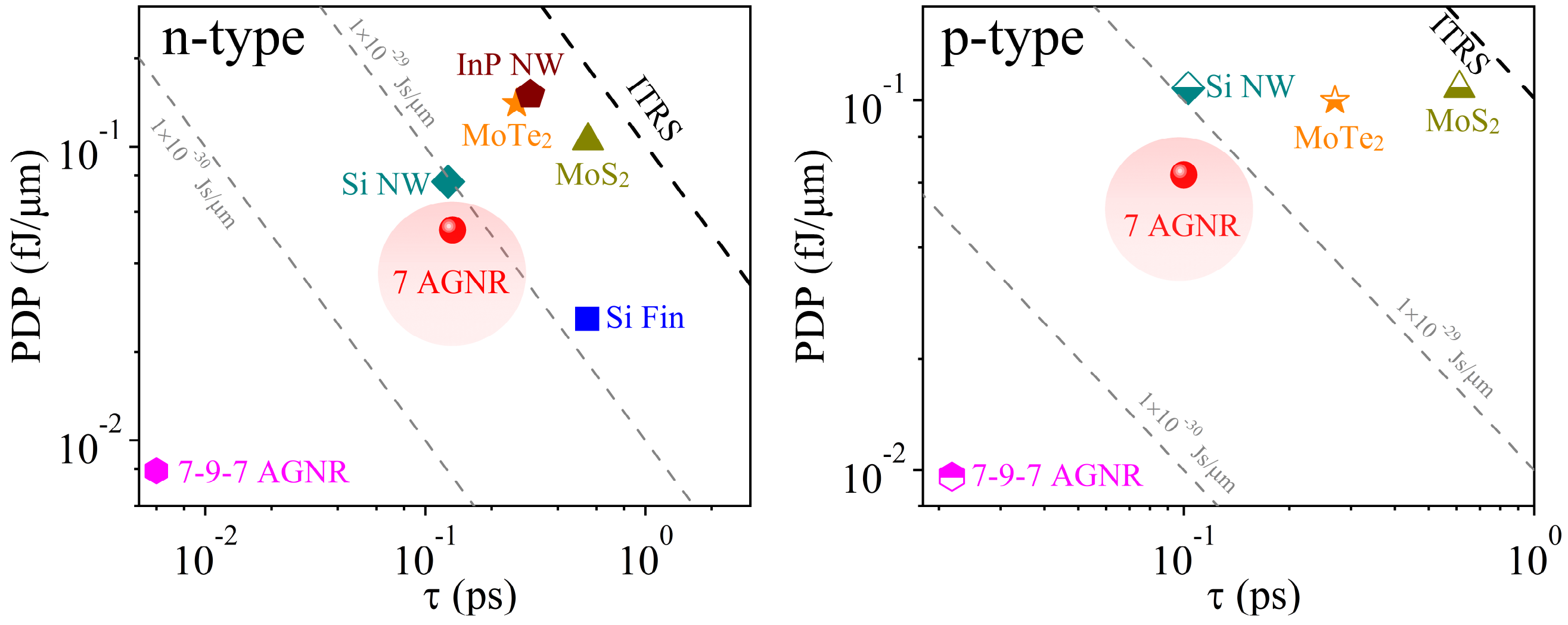

**Figure 5.** PDP versus $\tau$ in the low-dimensional material FETs at $L_g$ = 3 nm for the (a) *n*-type and (b) *p*-type devices. The black dashed lines indicate the ITRS requirements for the energy-delay product EDP = $\tau \times$ PDP. The data of the 7-9-7 AGNR, ML $MoS_2$, ML $MoTe_2$, Si Fin, Si NW, and InP NW FETs are calculated by the quantum transport simulations.[19, 35-37, 43, 44]

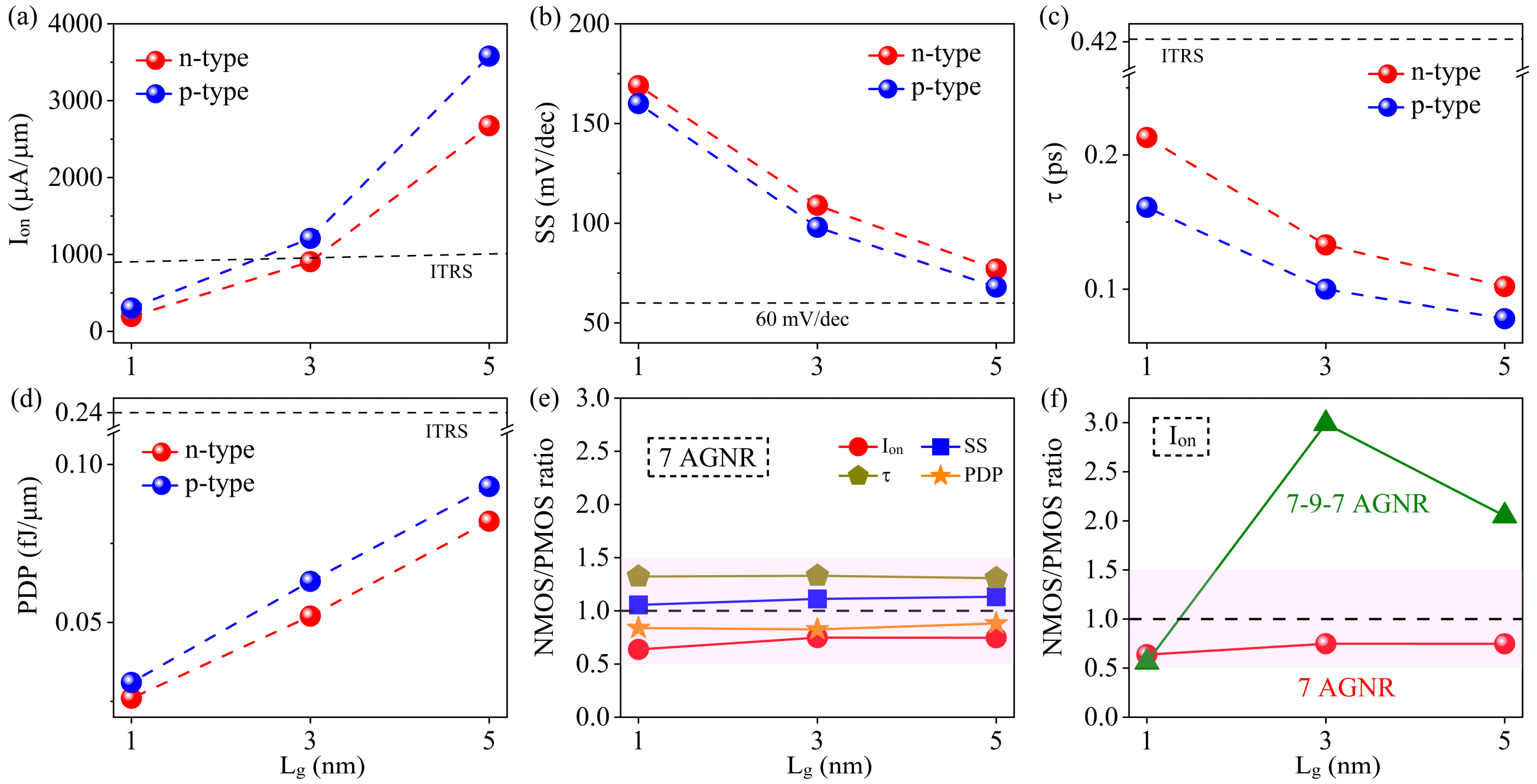

**Figure 6.** UL-optimized (a) $I_{\mathrm{on}}$, (b) *SS*, (c) $\tau$, and (d) PDP of the *n*-type and *p*-type 7 AGNR FETs at different gate lengths. (e) NMOS/PMOS ratios of the figures of merits ($I_{\mathrm{on}}$, *SS*, $\tau$, PDP) for 7 AGNR FETs. (f) Comparison of the $I_{\mathrm{on}}$ NMOS/PMOS ratio between 7 AGNR and 7-9-7 AGNR FETs.

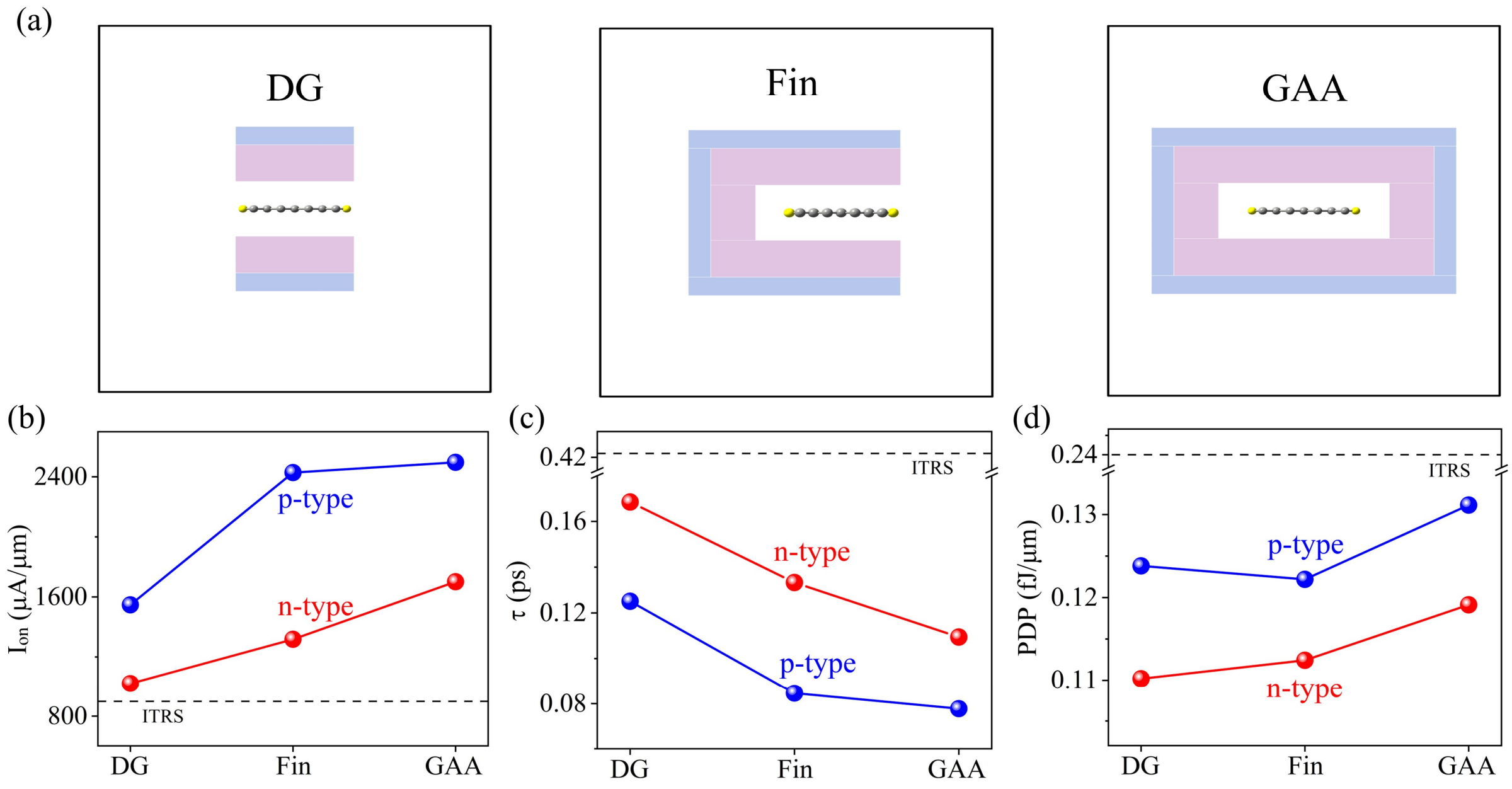

**Figure 7.** (a) Gate engineering of 7 AGNR FETs. (b) $I_{on}$, (c) $\tau$, and (d) PDP variation after gate engineering.

# TOC

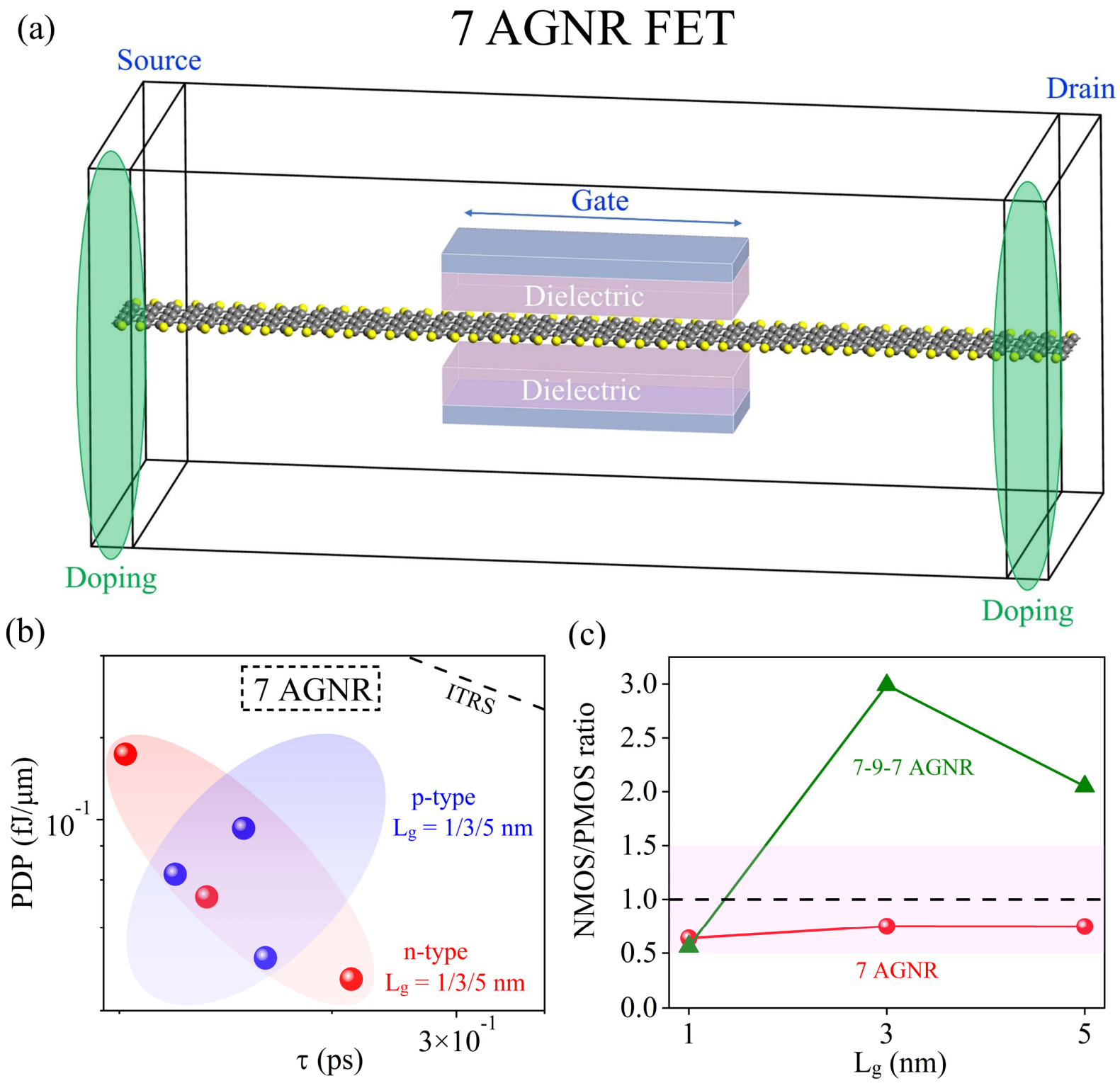